\documentstyle[prl,aps,epsfig]{revtex}
\begin{document}

\setlength{\unitlength}{1cm}
\begin{title}
{Normal state thermodynamics of cuprate superconductors }
\end{title} 
\author{A.S. Alexandrov}
\address{Loughborough University, Loughborough, Leics LE11 3TU (UK)}
\author{G.J. Kaye}
\address{IRC in Superconductivity, University of Cambridge, Madingley 
Road, 
Cambridge CB3 0HE (UK)}

\maketitle
\begin{abstract}
We propose a microscopic explanation of the  pseudogap features discovered in the normal state  
specific heat and magnetic susceptibility of 
cuprates. In the framework of the bipolaron theory of high-$T_{c}$ superconductors we explain the   
magnitude of the carrier specific heat and  
susceptibility as well as their universal scaling with temperature over a wide range of doping.
\end{abstract}

There is strong evidence for the normal state pseudogap in high-$T_{c}$ 
cuprates from magnetic susceptibility \cite{jef}, specific heat 
\cite{lor}, angle-resolved photoemission (ARPES) \cite{din}, 
tunnelling \cite{ren}, and some 
kinetic measurements \cite{bat}. One view supported by ARPES  is that the 
gap reflects precursor superconducting correlations in the BCS-like state 
below some  characteristic temperature $T^{*}$ without long range phase 
coherence \cite{eme}. Testing of this hypothesis with specific heat
\cite{lor} and 
tunnelling \cite{ren} data, it is found that this view cannot be sustained. In
particular,
there is no sign that the gap 
closes at a given temperature $T^{*}$, which
 rules out any role of superconducting phase or spin fluctuations \cite{ren}. On the other hand,  
the strong-coupling extension of the BCS 
theory based on the  multi-polaron perturbation technique  firmly predicts the transition to a  
charged Bose liquid in the crossover region  of  
the BCS coupling constant $\lambda\simeq 1$ \cite{ale}. 
 The (bi)polaronic theory of carriers in cuprates, confirmed by infrared spectroscopy  
\cite{tim} and by the isotope effect on the carrier mass 
\cite{mul},  provides a natural microscopic explanation of the normal state gap  \cite{alemot}.
   Within the framework of the theory, the ground state of cuprates 
   is a charged 
Bose-liquid of intersite bipolarons where single polarons exist only as excitations with an  
energy of $\Delta/2$ or more.  A characteristic 
temperature $T^{*}$ of the normal phase is a crossover temperature of the order of $\Delta/2$  
where the population of the upper polaronic band 
becomes comparable with the bipolaron density.  Along this line the normal state kinetics of  
cuprates has  been explained 
\cite{alebra,alekab} and a theory of tunnelling has been developed \cite{aletun} describing  
essential spectral features of STM and PCT 
conductance.
In this Letter, we find a universal temperature scaling of the specific heat and magnetic  
susceptibility of $YBa_{2}Cu_{3}O_{7-\delta}$ and 
provide a microscopic explanation with bipolarons and thermally excited polarons.   
The central ideas of our model are as follows
\begin{itemize}
\item Charge carriers are intersite real-space pairs of holes
 sitting 
on 
in-plane and  apical oxygens \cite{ale2}.
\item In addition to introducing hole charge carriers, doping also introduces considerable  
disorder and localised states. Owing to interparticle Coulomb repulsion \cite{ref}, a 
localisation 
well contains either a bipolaron or an unpaired polaron but not both. The interaction of polarons 
and bipolarons in the extended states is taken into account within the Hartree-Fock approximation  
and included in their band dispersion.  
\item At finite temperatures, a fraction of the carriers exist as unpaired hole polarons.
These particles are responsible for the magnetic response of the system. 
\end{itemize}
We also employ the simplification that the tunnelling probability between 
localisation wells is negligible. This allows the
partition function $Z_l$ for the localised part of the system to be 
written as
\begin{eqnarray}
Z_l &=& \prod_i Z_i \nonumber \\
Z_i &=& 1 + 2e^{(\mu-E_i)\beta} + e^{2(\mu-E_i+\Delta/2)\beta} ,
\label{eq:startpot}
\end{eqnarray}

where we have assumed the no double occupancy condition.  $\Delta,\ \mu$ and $E_{i}$ are 
respectively the bipolaron  
binding energy, chemical potential and a 
single-particle energy level of the well, whilst $\beta = 1/k_{B}T$. 
The point to note about equation (\ref{eq:startpot}) is that the localised partition function  
cannot be factorised into a product of two particle and 
one particle partition functions. The physics of localised bipolarons and polarons is thus {\em  
not separable} implying that only {\em one} 
density of states (DOS) profile should be taken for localised particles.
The  density of localised particles is determined by
\begin{equation}
 n_l  =  -\frac{\partial \Omega_l}{\partial \mu} 
\end{equation}
with $\Omega_l = - \beta^{-1} \log Z_l$. This gives
\begin{equation}
n_l = 2\int_{-\infty}^{0}  \rho_l (E) f_l (E) dE
\end{equation}
where
\begin{equation}
f_l (E)  =  \{1 + g[\beta (E-\mu-\Delta/2)]\}^{-1} 
\end{equation}
with
$g(\xi) =exp(\xi) cosh(\xi/2+\beta \Delta/4)/cosh(\xi/2-\beta \Delta/4)$. $\rho_l (E)$ refers  
to the density of localised states {\em per spin}.
We can then write for the 
number conservation condition:
\begin{equation}
2 n_b  + 2 n_p  + n_l = x,
\label{eq:number}
\end{equation}
where $n_{b,p}$ is the density of delocalised bipolarons and polarons, respectively and  $x$  the  
doping per unit cell. For $La_{2-x} Sr_{x} CuO_4$ $x$ is given by the atomic 
concentration of $Sr$ whilst in $YBa_{2}Cu_{3}O_{7- \delta}$, $x=2(1-\delta)/3$. 
The free particle  density is given by
\begin{eqnarray}
n_{b} &=& \int_{0}^{\infty} dE \rho_{b}(E) f_{b}(E) \nonumber \\
n_{p} &=& \int_{0}^{\infty}dE\rho_{p}(E)f_{p}(E)
\end{eqnarray}
where $f_{b}(E)=\{exp[\beta(E-2\mu-\Delta)]-1\}^{-1}$ and $f_{p}(E)=\{exp[\beta(E-\mu)]+1\}^{- 
1}$, so that equation (5)  allows us to determine 
the chemical potential $\mu(T)$ if bipolaronic and polaronic DOS, $\rho_{b,p}(E)$ are known.
  The finite bipolaron 
bandwidth, the one-dimensional singularity of (bi)polaronic DOS
\cite{ale2}, 
and a finite 
width of the localised  tail  give rise to a Shottky-like anomaly of the specific heat and a 
Curie- 
like temperature dependence of the susceptibility 
which are observed  at high temperatures in overdoped samples as explained in Ref. \cite{alekab}.
Here we consider the underdoped region, where  the 
potential wells are deep and 
impurity-scattering broadening of the Van-Hove singularities (VHS) large due to
 ineffective screening by carriers. The previous analysis \cite{alemot,aletun} 
showed that the characteristic width of the localised tails  and 
 VHS is above room 
 temperature in underdoped samples. We 
 can thus neglect any DOS structure for the relevant temperature range by taking $\rho_l  
(E)=\rho_{p}(E) = 2 \rho_{b}(E) \simeq N(0)$ with 
$N(0)$  a single-particle DOS at the mobility edge, $E=0$. The bipolaron chemical potential  
$2\mu +\Delta$ is then pinned at the mobility 
edge, giving  $\mu=-\Delta/2$, as follows from Eq.
(\ref{eq:number}) for $k_{B}TN(0) \ll 1$. This assumption greatly simplifies
further 
calculations. Including the contribution of
 delocalised bipolarons, thermally excited polarons and localised 
carriers we obtain 
the total energy as
\begin{equation}
E(T)=E_{0}+{N(0)\over{\beta^{2}}}\int_{0}^{\infty}d\xi \{\xi [coth(\xi)-1] 
+
(4\xi+\beta \Delta)[1-tanh(\xi + \Delta \beta /4)] + 2\xi [g(-\xi)^{-1}+1]^{-1}\}.
\end{equation}  
Here $E_{0}$ is a temperature independent (negative) constant.
As a result we find a universal temperature scaling of the energy, 
$E(T)=f(\beta \Delta)$, which allows us to extract the normal state gap
from 
 the experimental specific heat $C=\partial E/\partial T$ without 
any fitting parameters as shown in Fig.1. In the low-temperature limit,
$\beta\Delta \gg 1$ 
we get $g(\xi)\simeq exp(2\xi)$ and a linear specific heat with an 
exponential correction
\begin{equation}
C\simeq k_{B}N(0) \beta^{-1} 
\left[{\pi^{2}\over{4}}+{\beta^{2}\Delta^{2}\over{2}}exp(-\beta\Delta/2)\right].
\end{equation}
This result is in contrast with an expectation that the specific heat of nondegenerate   
bipolarons is temperature independent above $T_{c}$. 
The random 
potential as well as a low-dimensional DOS
 pins the chemical potential 
at the mobility edge even in the normal state, so the bipolaron density (and hence the specific  
heat) is proportional to temperature. The latter 
leads to a temperature dependent Hall effect\cite{alebra} and explains other anomalous kinetic  
properties of cuprates \cite{ref2}. Half of the 
bipolaron binding energy  $\Delta/2$, which is an energy gap between the bottoms of bipolaronic  
and polaronic bands has been estimated 
from $400 K$ to $50 K$ depending on doping \cite{alekab}.  In this temperature range one has to  
calculate $E(T)$ and $\gamma=C/T$ by 
numerical integration of Eq.(7) with the result shown in Fig.1. There is a clear scaling of  
experimental $\gamma$ with $\beta \Delta$ in a wide 
doping range of $YBa_{2}Cu_{3}O_{7-\delta}$. The corresponding values of $\Delta$ are shown in 
Fig.2. They follow the same doping
dependence as that determined phenomenologically \cite{lor} and are of the same order of 
magnitude. It should be noted though that the d-wave approach
taken by Loram {\it et al} gives consistently higher gap values than those found here. 
Nonetheless, d or s-wave like
gaps in the DOS can be obtained easily within this model by adjusting the nature of the particle-
particle interaction or the $k-$dependence
of the polaronic energy. Such discrepancies are not thus a significant problem.
A drop of $\gamma$ 
at higher temperatures (Fig.1)  is  due 
to a finite bipolaron bandwidth as 
discussed above.

Following Loram's analysis \cite{lorfr}, we compare $\gamma$ with the differential magnetic 
susceptibility $\chi^* = \partial(\chi T)/\partial T$.
The experimental data for $\chi^*(T)$ 
are perfectly consistent with
 our model. There are two contributions to 
the magnetic response, from 
delocalised (thermally excited) polarons, $\chi_{p}$, and from localised ones, $\chi_{l}$. For  
the first  contribution we obtain by the use of the 
Kubo formula for free fermion magnetisation, \begin{equation} \chi_{p}(T)= 
2\mu_{B}^{2}N(0)[exp(\beta \Delta/2)+1]^{- 
1}, \end{equation} where $\mu_{B}$ is the Bohr 
magneton. 
The single- well  partition function in  an external magnetic field, $H$
  is given by
\begin{equation}
Z_i = 1 + e^{2(\mu- E+\Delta/2)\beta} + e^{(\mu-E + \mu_B H)\beta} + 
e^{(\mu-E - \mu_B H)\beta}
\end{equation}
Differentiating twice the corresponding $\Omega$ potential with respect 
to the magnetic field yields 
\begin{equation}
\chi_l (T) =\mu_{B}^{2}\beta  \int_{-\infty}^{0} dE\rho_l (E)f_{l}^{p}
(E),
\end{equation}
where
$f_{l}^{p} (E) = [1 + exp(\beta \Delta/2) cosh((E-\mu-\Delta/2)
\beta)]^{-1}$ 
is the distribution function of localised 
polarons. If DOS is a constant $\rho_l (E)=N(0)$, and temperature is low, $\beta \Delta \gg 1$ we  
obtain an exponential temperature 
dependence of the spin susceptibility as \begin{equation} \chi(T)=\chi_{p}(T)+\chi_{l}(T) \simeq  
2\mu_{B}^{2}N(0)(1+\pi/4) exp(-\beta \Delta/2).
\end{equation}
The numerical integration of Eq. (11) for the entire temperature range with the constant DOS  
yields a universal scaling of $\chi^*$ as a function 
of $\beta \Delta$. This is nicely confirmed by experiment, as shown in Fig.3. It is remarkable, 
that  
with about the same $\Delta$ and DOS (see Fig.2) 
one can describe both the specific heat and spin susceptibility of underdoped  
$YBa_{2}Cu_{3}O_{7-\delta}$.  This is at variance 
with some opinions that the experimental Wilson ratio is difficult to understand within the 
framework  
of our model. In fact, thermally excited 
polarons provide the spin susceptibility and a finite Wilson ratio close to the experimental one,  
while the binding energy of bipolarons is 
responsible for the normal state 'gap'.
We may therefore conclude that 
 the formation of real space pairs (bipolarons) above $T_{c}$ and their
partial 
 localisation by the random potential are
essential features in  
describing the normal state thermodynamics of 
$YBa_{2}Cu_{3}O_{7-\delta}$ and other cuprates exhibiting similar normal state gap.
 The bipolaron theory can explain such non-Fermi liquid features as a large carrier entropy,  
the gap above $T_{c}$, temperature dependence of  $\gamma$ and $\chi$ and their ratio. Another 
strong indication of the existence of bipolarons comes from the  
resistive and thermodynamic measurements 
in the critical region. A divergent upper critical field was measured in many cuprates as  
predicted by one of us \cite{aleH}, and the magnetic 
field dependence of the specific heat jump is just that of the charged Bose-gas \cite{alebee}.
 We greatly appreciate the enlightening discussions with A. Blackstead, J.R. Cooper, J.T. 
Devreese, J.D. Dow,  
A. Junod, H. Kamimura, W.Y. Liang, J. Loram, J.L. Tallon and G. Zhao.

\centerline{{\bf Figure Captions}}
Fig.1 Universal scaling of $\gamma/k_B^2 N(0)$  with $2 k_{B}T/\Delta$ compared with theory 
(line) for  
$YBa_{2}Cu_{3}O_{7-\delta}$ (N(0)= $1.17 \mathrm{eV}^{-1}$ per spin).
Fig.2  Theoretical normal state gap as a  
function of doping.
Fig.3 Universal scaling of the differential spin susceptibility, $\chi^*(T)/\mu_B^2 N(0) = 
(\chi^*_{exp} - 0.39 \times 10^{-4} {\mathrm{emu/mole}})/\mu_B^2 N(0)$ compared with  
theory (line). \\
For experimental data, see Loram {\it et al} \cite{lorfr}

\begin{figure}
\begin{picture}(16,20)
\put(0,18){\parbox{16cm}{ \bf Figure 1: Universal scaling of $\gamma/k_B^2 N(0)$  with $2 
k_{B}T/\Delta$ compared with theory (line) for  
$YBa_{2}Cu_{3}O_{7-\delta}$ (N(0)= $1.17 \mathrm{eV}^{-1}$ per spin).}}
\put(0,4){\epsfig {file = 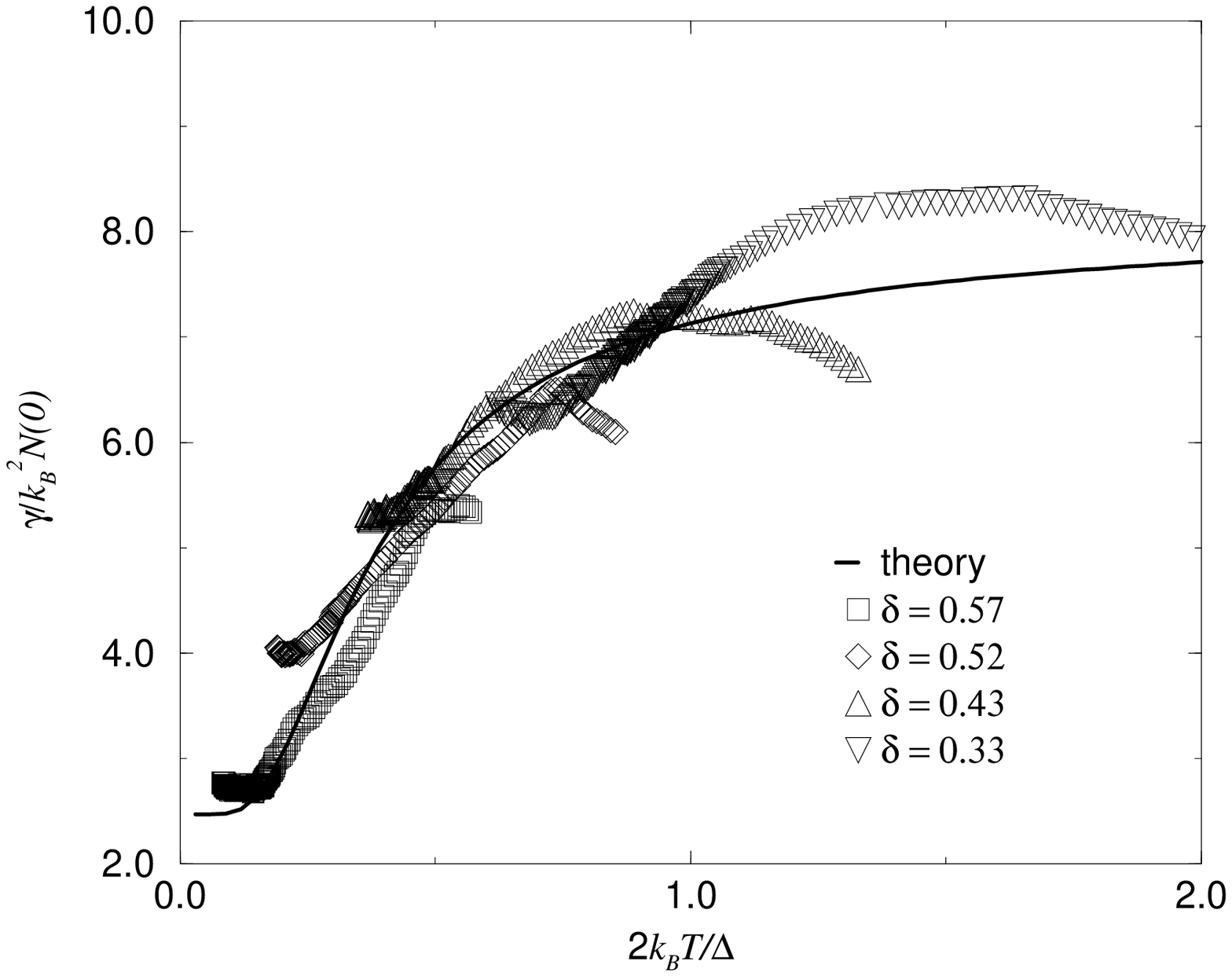, height = 15cm, clip = true, bbllx = 6mm, bblly = 0cm,
bburx = 26cm, bbury = 18cm }}
\end{picture}
\end{figure}

\pagebreak

\begin{figure}
\begin{picture}(16,24)
\put(0,23){\parbox{16cm}{ \bf Figure 2: Theoretical normal state gap as a  
function of doping.}}
\put(0,8){\epsfig {file = 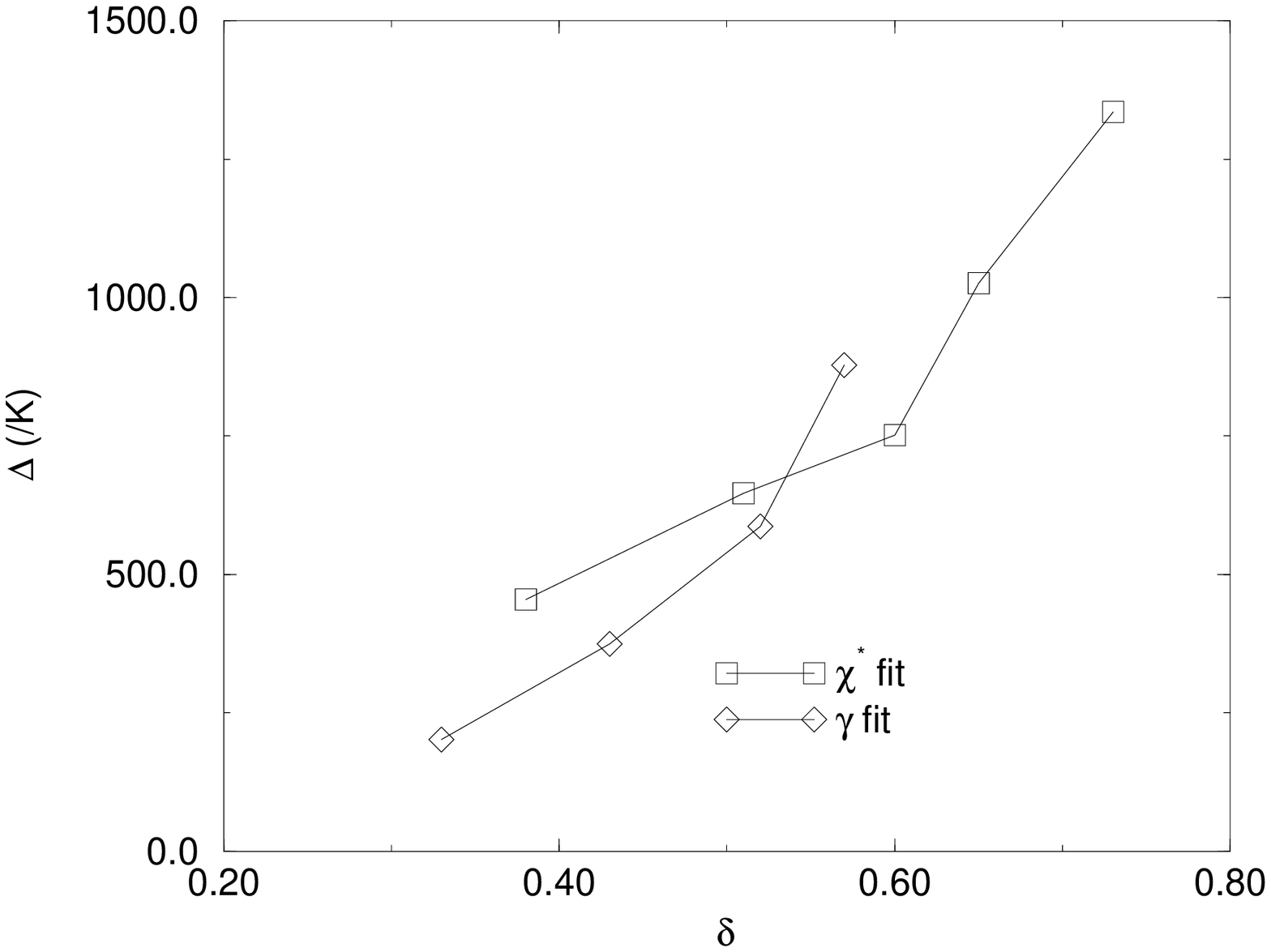, height = 15cm, clip = true, bbllx = 6mm, bblly = 0cm,
bburx = 26cm, bbury = 18cm }}
\end{picture}
\end{figure}

\pagebreak

\begin{figure}
\begin{picture}(16,24)
\put(0,23){\parbox{16cm}{ \bf Figure 3: Universal scaling of the differential spin 
susceptibility, $\chi^*(T)/\mu_B^2 N(0) = (\chi^*_{exp} - 0.39 \times 10^{-4} 
{\mathrm{emu/mole}})/\mu_B^2 N(0)$ compared with  
theory (line).}}
\put(0,8){\epsfig {file = 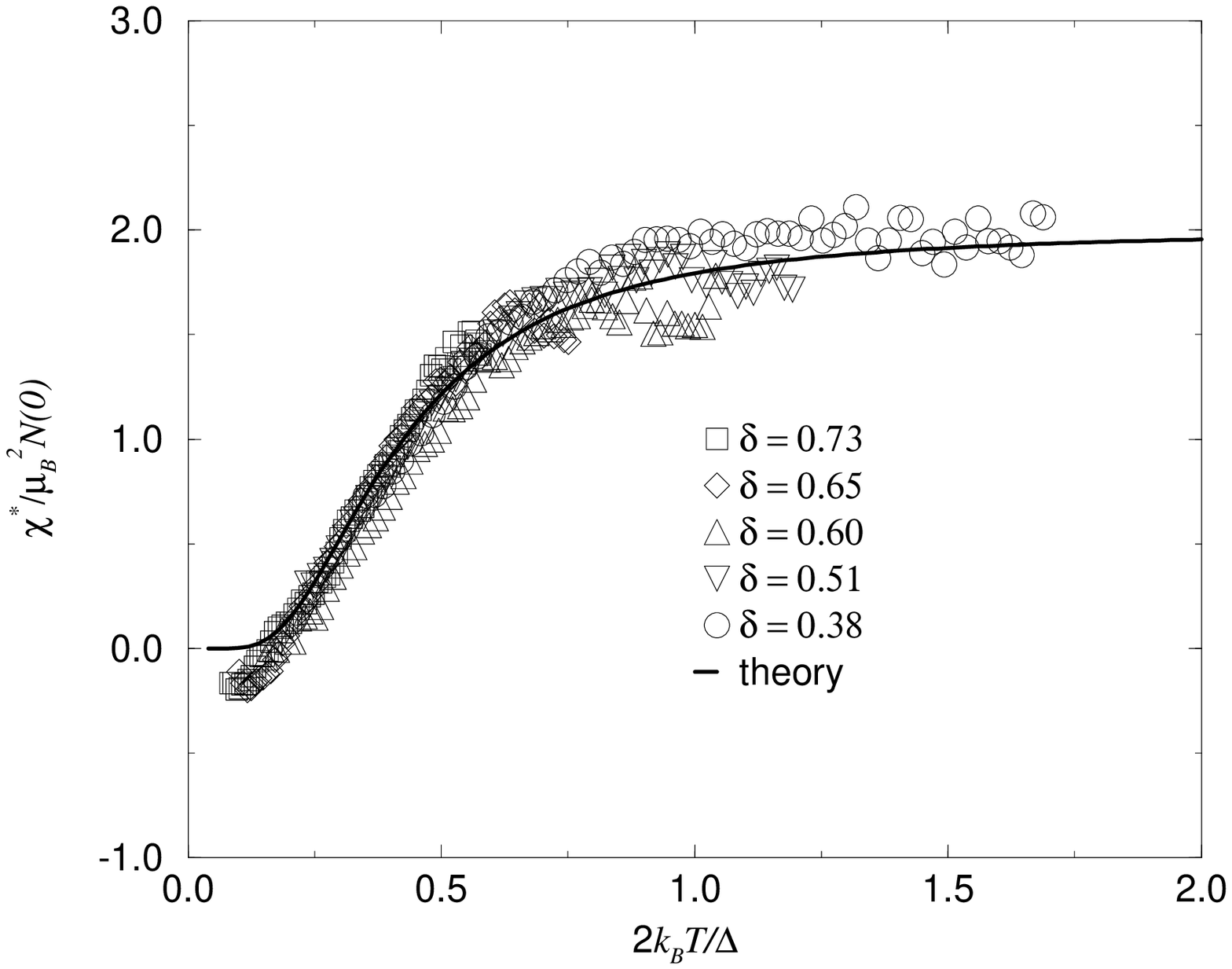, height = 15cm, clip = true, bbllx = 6mm, bblly = 0cm,
bburx = 26cm, bbury = 18cm }}
\end{picture}
\end{figure}


\begin{thebibliography}{99}
\bibitem{jef}
D.C.  Johnston, Phys. Rev. Lett. {\bf 62} 957 (1989)
\bibitem{lor}
J.W. Loram $et$ $al$,
 J. of Superconductivity {\bf 7}, 243 (1994)
\bibitem{din}
Z.-X. Shen and J.R. Schrieffer, Phys. Rev. Lett. ${\bf 78}$, 1771 (1997), and references therein.
\bibitem{ren}
Ch. Renner $et$ $al$, Phys. Rev. Lett. ${\bf 80}$, 149 (1998).
\bibitem{bat}
H.W. Hwang $et$ $al$, Phys. Rev. Lett. ${\bf 72}$, 2636 (1994).  \bibitem{eme} V.J. Emery, S.A.  
Kivelson, Nature {\bf 374}, 434 (1995);
V.J. Emery, S.A. Kivelson, and O. Zachar, Phys. Rev. B${\bf 56}$, 6120 (1997).
\bibitem{ale} A.S. Alexandrov,  Zh.Fiz.Khim. ${\bf 57}$, 273 (1983) (Russ.J.Phys.Chem.${\bf 57}$,  
167 (1983));  Phys. Rev. B${\bf 46}$, 2838 
(1992).  
\bibitem{tim}  D.B. Tanner and T. Timusk, in 'Physical Properties of High-Temperature  
Superconductors III', ed. D.M.  Ginsberg, World 
Scientific, Singapore (1992);
E.K.H. Salje, A.S. Alexandrov, and W.Y. Liang (eds),  'Polarons and Bipolarons in High-$T_{c}$  
Superconductors and Related Materials', 
Cambridge University Press, Cambridge (1995); 
 J.T. Devreese and J. Tempere, to be published in Solid State Commun. 
 (1998).
\bibitem{mul}
G. Zhao, M.B. Hunt,H.  Keller, and K.A. M\"uller, Nature ${\bf 385}$, 236 (1997).
\bibitem{alemot}
A.S. Alexandrov and N.F. Mott, Rep. Prog. Phys. ${\bf 57}$ 1197 (1994);
{\it Polarons and Bipolarons}, 
World-Scientific (1995). 
\bibitem{alebra}
A.S. Alexandrov, A.M. Bratkovsky and N.F. Mott, Phys. Rev. Lett. {\bf 72}, 1734 (1994)
\bibitem{alekab}
A.S. Alexandrov, V.V. Kabanov and N.F. Mott, Phys. Rev. Lett. ${\bf 77}$, 4796 (1996).
\bibitem{aletun}
A.S. Alexandrov, unpublished (1998).
\bibitem{ale2}
A.S. Alexandrov,  Phys.Rev. B$ {\bf 53}$, 2863 (1996); there is strong experimental evidence for  
the important role played by apical oxygen holes 
from site-specific x-ray absorption (M. Merz $et$ $al$, to be published (1998)) and the  
site-specific effect of $Ba$ and $Y$ 
substitution for $Pr$ in $YBa_{2}Cu_{3}O_{7}$ (J.D. Dow, U. Howard and A. Blackstead, Bulletin of  
the American Physical Society, ${\bf 43}$, 
877 (1998)).
\bibitem{ref} Bipolaron-bipolaron and bipolaron-polaron interaction is a long-range Coulomb  
repulsion  in the case of dispersionless phonons 
\cite{alemot}. 
\bibitem{ref2}
Both localised and 
extended carriers contribute to the mid-infrared conductivity, which is incoherent. This 
incoherence is also
seen in  ARPES and 
tunnelling spectra. Hence, any Drude-like fit to the optical conductivity is misleading and  
cannot provide a true temperature dependence of the 
carrier density.
\bibitem{lorfr} J.R.  Cooper, J.W.  Loram, J.  Phys.  I France {\bf 6}, 2237 (1996)

\bibitem{aleH}
A.S. Alexandrov, Phys. Rev. B${\bf 48}$, 10571 (1993); there were some doubts concerning the  
applicability of this result to cuprates (A.A.  
Abrikosov, Phys. Rev. B {\bf 56}, 446 (1997) ) based on an (incorrect)  assumption that 
bipolarons could be  
formed only at a very large BCS coupling 
constant $\lambda >(M/m_{e})^{1/2}$ and that there was no 'pseudogap' effect in overdoped 
cuprates  
with the divergent upper critical field. 
However, it is well known \cite{alemot}, that the polaronic collapse  has nothing to do with the  
adiabatic parameter $M/m_{e}$, so bipolarons 
appear at $\lambda \simeq 1$, and the pseudogap $is$ found in overdoped samples also  
\cite{ren}.
\bibitem{alebee}
A.S. Alexandrov $et$ $al$, Phys. Rev. Lett. ${\bf 79}$, 1551 (1997).
\end{thebibliography}
\end{document}